# Cultural evolution in Vietnam's early 20th century: a Bayesian networks analysis of Franco-Chinese house designs

Working Paper No. PKA-1901

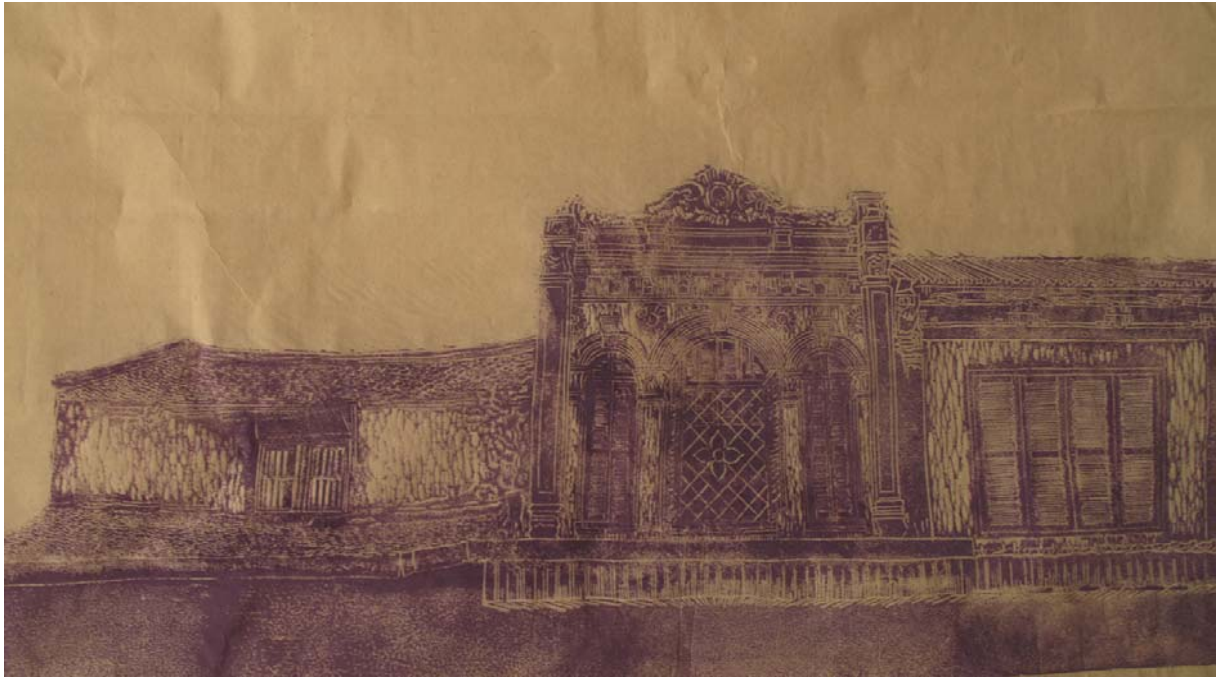

©2016 Bui Quang Khiem

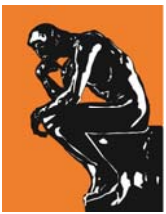

*A.I. for Social Data Lab*

**Tech:** QH Vuong  0000-0003-0790-1576; VP La  0000-0002-4301-9292
**Research team:** HM Tung  0000-0002-4432-9081; HKT Nguyen  0000-0003-4075-9823; TT Vuong  0000-0002-7262-9671; HM Toan  0000-0002-8292-0120
**Admin:** Thu-Ha Dam (Vuong & Associates, Hanoi, Vietnam)
**Email:** qvuong.ulb@gmail.com

# Cultural evolution in Vietnam's early 20th century: a Bayesian networks analysis of Franco-Chinese house designs

Quan-Hoang Vuong
Quang-Khiem Bui
Viet-Phuong La
Thu-Trang Vuong
Manh-Toan Ho
Hong-Kong T. Nguyen
Hong-Ngoc Nguyen
Kien-Cuong P. Nghiem
Manh-Tung Ho

# Cultural evolution in Vietnam's early 20th century: a Bayesian networks analysis of Franco-Chinese house designs

Quan-Hoang Vuong[1,2] orcid, Quang-Khiem Bui[3], Viet-Phuong La[1,2], Thu-Trang Vuong[5], Manh-Toan Ho[1,2,4], Hong-Kong T. Nguyen[4] orcid, Hong-Ngoc Nguyen[6], Kien-Cuong P. Nghiem[7], Manh-Tung Ho[1,2,8,9] * orcid

[1] Centre for Interdisciplinary Social Research, Phenikaa University, Yen Nghia Ward, Ha Dong District, Hanoi, 100803, Vietnam

[2] Faculty of Economics and Finance, Phenikaa University, Yen Nghia Ward, Ha Dong District, Hanoi 100803, Vietnam

[3] Hanoi College of Arts, 7 Hai Ba Trung Street, Hoan Kiem District, Hanoi 100000, Vietnam

[4] A.I. for Social Data Lab, Vuong & Associates, 3/161 Thinh Quang, Dong Da District, Hanoi 100000, Vietnam

[5] Sciences Po Paris, Campus de Dijon, 21000 Dijon, France

[6] Ho Chi Minh City University of Fine Arts, Ho Chi Minh City, Vietnam

[7] Vietnam-Germany Hospital, 16 Phu Doan street, Hoan Kiem district, Hanoi 100000, Vietnam

[8] Vietnam Academy of Social Sciences, Institute of Philosophy, 59 Lang Ha Street, Ba Dinh District, Hanoi 100000, Vietnam.

[9] Graduate School of Asia Pacific Studies, Ritsumeikan Asia Pacific University, Oita Prefecture, 874-8577, Japan.

*(\*) Corresponding author; email: tung.homanh@phenikaa-uni.edu.vn*

**Abstract**

The study of cultural evolution has taken on an increasingly interdisciplinary and diverse approach in explicating phenomena of cultural transmission and adoptions. Inspired by this computational movement, this study uses Bayesian networks analysis, combining both the frequentist and the Hamiltonian Markov chain Monte Carlo (MCMC) approach, to investigate the highly representative elements in the cultural evolution of a Vietnamese city's architecture in the early 20th century. With a focus on the façade design of 68 old houses in Hanoi's Old Quarter (based on 78 data lines extracted from 248 photos), the study argues that it is plausible to look at the aesthetics, architecture and designs of the house façade to find traces of cultural evolution in Vietnam, which went through more than six decades of French colonization and centuries of sociocultural influence from China. The in-depth technical analysis, though refuting the presumed model on the probabilistic dependency among the variables, yields several results, the most notable of which is the strong influence of Buddhism over the decorations of the house façade. Particularly, in the top 5 networks with the best Bayesian Information criterion (BIC) scores and $p<0.05$, the variable for decorations (DC) always has a direct probabilistic dependency on the variable B for Buddhism. The paper then checks the robustness of these models using Hamiltonian MCMC method and find the posterior distributions of the models' coefficients all satisfy the technical requirement. Finally, this study suggests integrating Bayesian statistics in social sciences in general and for study of cultural evolution and architectural transformation in particular.

# Introduction

> "There is no better illustration of the idea that while the past may live in the present, it is the present that constructs the past and is constructed in the past."
>
> – *The Country of Memory* by Hue-Tam Ho Tai (2001)

The old streets and houses of Hanoi have inspired generations of local artists, who with their paints and canvases have in turn immortalized the crimson-brick roofs, the faded gray walls, the whitewashed concrete electric poles, the faceless ladies in conical hats, and the nearly empty winding streets and alleys, to name a few notable features of this genre. These images are most often reminiscent of the works by the artist Bui Xuan Phai, whose name is now synonymous with oil paintings about bygone Hanoi Old Quarter—"*Phố Phái*" (literally, "Street Phai") (Taylor, 1999). Bui Xuan Phai (1920-1988) is one of the "Four Pillars" of Vietnamese paintings, belonging to the realm of folklore and myth (Naziree, 2006; Taylor, 1999; Thai, 1994). "*Phố Phái,*" originated from the early 1960s (TT&VH, 2010), implies not just the Vietnamese respect for a great artist but also a symbol of aesthetics in the Vietnamese souls. It conjures up memories of the second half of the twentieth century – a time when wars ravaged Vietnam, when the revolutionary spirit had at first appealed to the cultural intellectuals but eventually disillusioned some because of its attempt to supplant the artists' romantic expressionism with socialist realism, to ultimately create art for the sake of politics (Naziree, 2006). "*Phố Phái*" thus is a reminder of the angst and poverty of wartime Hanoi as much as of a modernization strategy that exploits French colonial architecture. Perhaps this is what the Vietnam studies scholar Tai (2001) refers to when he writes about the "longings for the trappings of modernity" as expressed in pictures of old streets, all against the face of high-rise construction. Seeking to understand the cultural evolution to modern Vietnam during the latter half of the twentieth century, this paper will analyze the façade designs of old houses around Hanoi, upon which questions about nostalgia, traditions and progress toward modernity will be discussed.

**Figure 1:** A painting of Hanoi Street by Bui Xuan Phai (2012)

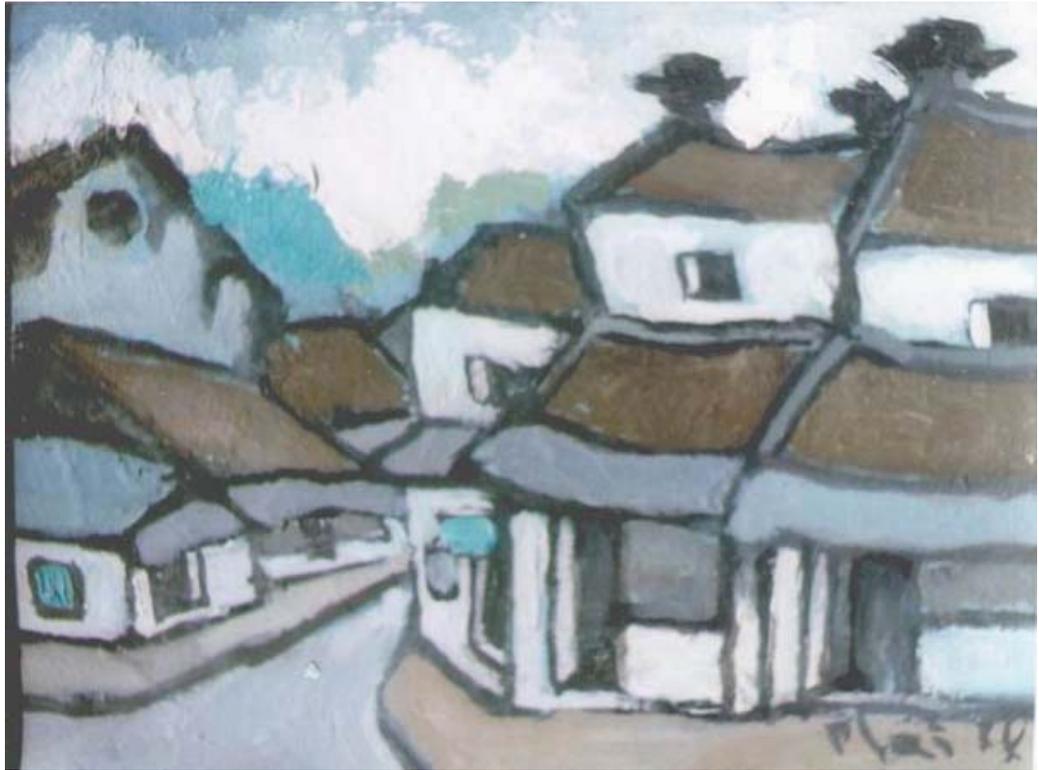

To understand this cultural evolution and its implications, the paper proposes looking into why "*Phố Phái*" looks the way it does, for it is crucial to situate the paintings, and later Hanoi architecture, in the right historical context. The master behind it, Bui Xuan Phai, was educated at the École Supérieure des Beaux-Arts de l'Indochine *(Indochina College of Fine Arts)*, the former name of Vietnam University of Fine Arts (André-Pallois, 2016).  joined the cultural front to produce propaganda poster art in 1945, relocated to the countryside in 1958 for manual labor with the peasants under a Party policy, but by 1966 had decided to pursue artistic expressionism free from any political agency (Bui & Tran, 1998; Naziree, 2006). He retreated to his home in Hanoi where he was so poor, he lacked even the most basic instruments for painting. Yet, his resilience and love for the city had translated into his works—into the houses that may appear wobbly but have stood there for decades. In his paintings, the streets of Hanoi came alive in simple brushstrokes of bold borders and mundane gray and white colors. The cityscape in Phai's paintings, often understood in three periods of 1960-1970, 1970-1980, and 1980-1988, appeared in a familiar and simple form with subtle signs of evolution: the two-storied houses, attached to one another like stacking matchboxes with the same dark brown brick roofs, were old and undecorated, their doors perpetually shut; then as the 1970s-1980s came, the houses' shuttered windows were painted in more details instead of in one black brushstroke and the first sign of modernity emerged in the form of a lamppost.

These paintings have captured the very essence of what is known in Vietnam as the "tube house" ("*nhà ống*")—traditionally attached street houses whose widths are narrow while their lengths are very long (Kien, 2008a, 2008b). This research, inspired by "*Phố Phái*" and the cultural-historical continuity as reflected in Hanoi architecture, will delve into the elements of cultural and religious influences in the house designs, especially the façade. The rapid urbanization and commercialization in Vietnam mean that many of these old houses are being changed and may not stand the test of time. The internal structure of the houses would make for an interesting inquiry, but the research's focus on the house façade alone is driven by the principal concern over "face" in Confucian society. In other words, the front of house reflects the face of the family, and therefore, its culture. As Vuong, La, et al. (2018) have shown, the "cultural additivity" in Vietnamese architecture is reflected in the front of a house in the co-existence of French-styled columns, Confucian scrolls, the Taoist yin-yang sign, and the Buddhist lotus sculpture. This study will investigate the highly representative elements in the cultural evolution of Hanoi's architecture in the early 20$^{th}$ century using Bayesian networks analysis from both the frequentist and the Hamiltonian Markov chain Monte Carlo (MCMC) approach.

**Research Questions**

This research sets out to answer the following two questions:

**RQ1:** Is it plausible to look at the aesthetics, architecture and designs of the house façade in Hanoi to find traces of cultural evolution in the early 20$^{th}$ century in Vietnam, which went through more than six decades of French colonization and centuries of sociocultural influence from China?

**RQ 2:** What are the most notable elements that affect the perception on cultural evolution of Vietnamese people who are familiar with or have been exposed regularly to this type of architecture and cultural behavior?

**Literature Review**

This section will review the voluminous works on cultural evolution, identify relevant research from the literature, and provide an overview of the Vietnamese aesthetics and architecture.

*Research on cultural evolution*

In one of the early works on this topic, Boyd and Richerson (1996) have made the case for why culture and cultural variation are common in nature, but "cumulative cultural evolution" is rare because it requires the complex capacity for observational learning. The evolution of culture in human beings, manifested in the cognitive details of human social learning and inference, is also referred to as adaptive cultural processes that may result in both useful and maladaptive losses (Henrich, 2004; Henrich & McElreath, 2003). These ideas represent the line of argument that takes evolutionary biology as a reference point to illustrate the Darwinian evolutionary properties of human culture, such as "the selective retention of favorable culturally transmitted variants, as well as a variety of non-selective processes, such as drift" (Boyd & Richerson, 1985; Cavalli-Sforza & Feldman, 1981; Mesoudi, Whiten, & Laland, 2006; Youngblood & Lahti, 2018). This approach has become prominent over the erroneous view of fixed stages of cultural progress, mainly thanks to the pioneering works of Cavalli-Sforza and Feldman (1981) and Boyd and Richerson (1985). As Mesoudi et al. (2006) argue, the evolutionary framework, which applies mathematics and statistical analysis in modeling cultural changes, has rich potentials for phylogenetic analyses or population genetic models. Given that a more rigorous science of culture could improve the empirical investigation as well as applications of emerging knowledge, researchers in this field have been calling for theoretical unification (Claidière & André, 2012; Mesoudi et al., 2006).

Critiques of this viewpoint, however, argue that not only does it overlook the differences between intelligent, mindful evolution and oblivious evolution, but it also underestimates the independence of cultural traits from genetic fitness (Dennett & McKay, 2006). In particular, when writing about cultural evolution, one needs not presume that all cultural traits that do evolve will be fitness-enhancing (Dennett & McKay, 2006). Some scholars who favor theories on memetic evolution—which posits that culture is a "by-product of the evolved capacity for imitation that then took off on its own evolutionary trajectory"—also disagree with the idea that culture is an adaptation (Blackmore, 2006). There are additional concerns about lumping all cultural traits that belong to distinct phylogenies and have different structural components under the same lineages (Fuentes, 2006; Mulder, McElreath, & Schroeder, 2006), or the lack of causal relevance between "chances of survival" and "phenotypical consequences of cultural traits" (Borsboom, 2006). Overall common critiques suggest not simplifying modes of cultural transmission by use of analogy with modes of genetic transmission because the two domains do not share the same properties (Claidière & André, 2012; Fuentes, 2006; Mulder et al., 2006).

While the debates may go on over the nature of cultural evolution and its comparability to biological evolution, it has become clear that the study itself is an interdisciplinary field for bringing together researchers from a wide range of fields such as evolutionary biology, anthropology, psychology, sociology, and computer science (Youngblood & Lahti, 2018). According to a bibliometric analysis of research on this interdisciplinary field, there are seven major groups of subjects, namely (i) biological anthropology and archeology, (ii) mathematical modeling and dual-inheritance theory, (iii) cognitive linguistics and experimental cultural evolution, (iv) cross-cultural and phylogenetic studies, (v) computational biology and cultural niche construction, (vi) evolutionary psychology, and (vii) behavioral ecology and birdsong (Youngblood & Lahti, 2018).

In the extant literature, with regards to the evolution of architecture, scholars have looked at the rural cemetery of the Anglo-American (Schuyler, 1984), the Soviet structures under Stalin (Paperny, 2002), the distribution of Egyptian military bases over time and across borders (Ellen, 2004), the wooden long-houses on the Pacific northwest coast (Jordan & O'Neill, 2010), or the cultural function in certain architecture (Houston, 1998), to name a few. Large-scale architecture in particular has indeed captivated many scholars and archeologists due to their conspicuous scale, structural complexity and aesthetic value (Abrams, 1989). This sentiment is echoed in various studies; a study of classic Mayan buildings and remains highlights the power of architecture to stir up "romantic nostalgia for a lost world in which one has not participated, but which might be imagined or scientifically resurrected" (Houston, 1998). Along this vein, in examining the evolving house front designs in the old streets of Hanoi, this study also seeks to evoke a meaningful discussion on the intertwined relationship between a longing for the past and a yearning for modernization.

*Bayesian approach in the study of cultural evolution*

The emerging trends in the field of cultural evolution are the interdisciplinary and diversity of methodology (Mesoudi, 2017; Mesoudi et al., 2006; Youngblood & Lahti, 2018). There appears to be more emphasis on mathematical modeling or phylogenetic methods in order to increase the scientific rigor in the study of culture and anthropology (Easterbrook, 2014; Holden, Meade, & Pagel, 2005; Mace & Holden, 2005; Matthews, Tehrani, Jordan, Collard, & Nunn, 2011; Mesoudi, 2017; Youngblood & Lahti, 2018). In studying cultural evolution, one expects to learn about the continuity in cultural transmission and adoptions, and thus, the more diversified the methods for these studies are, the more insights would one gain from the endeavors. Inspired by this computational movement and keeping in mind the diversities of cultures worldwide, this paper takes a novel perspective and method in studying the cultural evolution in Vietnam. Using Bayesian networks analysis combining both the frequentist and the Hamiltonian Markov chain Monte Carlo (MCMC) approach, this study will investigate the important elements in the cultural evolution of Hanoi's architecture in the early $20^{th}$ century.

Some examples of Bayesian inference and models in the study of culture are: exploring the evolution of social learning (Perreault, Moya, & Boyd, 2012), analyzing the language evolution and cultural transmission through iterated learning (Dediu, 2009; Griffiths & Kalish, 2007; Reali & Griffiths Thomas, 2010), or in estimating the correlated evolution across cultures, such as the wealth transfer in different marriage systems (Pagel & Meade, 2005) and the evolution of the mating system (Pagel & Meade, 2006). The Approximate Bayesian Computation (ABC), which "enables the evaluation of multiple competing evolutionary models formulated as computer simulations," has been applied in archeology to find patterns of cultural evolution based on artefact frequencies (Crema, Edinborough, Kerig, & Shennan, 2014). In other notable investigations, scholars have used Bayesian phylogenetic analysis to identify incongruent descent histories in Iranian tribal textile traditions (Matthews et al., 2011), built Bayesian radiocarbon models to estimate the timing of the cultural transition in southern Scandinavia over 11,000 years ago (Riede & Edinborough, 2012) or in eastern Fennoscandia (Oinonen et al., 2014; Pesonen, Oinonen, Carpelan, & Onkamo, 2012).

Meanwhile, Bayesian-based research on cultural evolution in Asia have looked at the agricultural origin of Japonic languages (Lee & Hasegawa, 2011), the revised chronology of the lower Yangtze in eastern China (Long & Taylor, 2015), the evolutionary history of Indian populations through a certain mitochondrial lineage (Kumar et al., 2008), the genetic and linguistic histories in Central Asia (Thouzeau, Mennecier, Verdu, & Austerlitz, 2017), or created a revolutionary chronological framework for prehistoric Southeast Asia based on radiocarbon modelling from a village in central Thailand (Higham & Higham, 2009).

The extant literature shows the rich potential of the Bayesian method in explicating historical and cultural phenomena around the world, even from very ancient time. Within the scholarship, however, there is a dearth of research on cultural evolution in Vietnam, especially those that use Bayesian statistics. This study hopes to fill in the gap through its novel approach in analyzing house façade designs in Hanoi. The following subsection will provide the background on Vietnamese architecture in general and Hanoi's Old Quarter in particular.

### *Research on the Franco-Chinese aesthetics and architecture in Vietnam*

To understand the shape and form of Hanoi today, one needs to be familiar with the roots of French influence in Vietnam. French troops captured the citadel of Gia Dinh in 1859, defeating the last Vietnamese dynasty Nguyen and paving the way for its establishment of French Indochina (initially comprising Annam, Tonkin, and Cochinchina) in 1887. Under French imperialism, the Cornudet Law that was passed by its government in 1919 stipulated the rules for urban planning and development in its colonies, such that contemporary Western construction techniques would take into account the native aesthetics and tropical humid climate (Le, 2013; Nguyen, 2014; Vongvilay, Shin, Kang, Kim, & Choi, 2015). For Indochina, Governor-General Maurice Long set up the Hanoi urban planning department, headed by the famed French architect Ernest Hebrard. The essence of Hebrard's architectural style, which becomes known as the Indochine style, lies in its effortless fusion of traditional Vietnamese and grand European elements for both aesthetics and practical purposes (Le, 2013). According to Nguyen (2014), the development of Indochinese architecture peaked in Vietnam in 1920-1945 and waned in the 1960s. In the post-1945 period, the designs fell into two categories: on the one hand, some projects were built in the name of traditions and filled with nostalgia and restoration of the past; on the other hand, there were architects looking for a more creative style in Vietnam, fitting in with the trend of internationalization at the time (Nguyen, 2014). In a slightly different account, Herbelin (2016) argues that this movement took place in the 1920-1930, with one wave of affluent Vietnamese people building houses entirely in the footsteps of the French and another pursuing a more modern style incorporating northern Vietnamese decorative elements. However, the architectural fusion, which used both indigenous construction materials such as bamboo, wood, hut, mud and metropolitan materials such as iron, concrete, tiles, and bricks, turned out to be costly and only those very wealthy could afford it (Herbelin, 2016).

In the past decade, there have been many monumental works on the French architecture and urban planning in Vietnam, particularly the French architecture of houses in Hanoi (*L'architecture française des maisons de Hanoï*) and Saigon (now Ho Chi Minh City). For example, Tran and Nguyen (2012) have provided a meticulous and comprehensive work on the legacy of French architecture in Hanoi, documenting the changing cityscape from as early as the mid-nineteenth century and comparing it with the construction of Indochine houses in Saigon in the early twentieth century. Walker (2011), together with photographer Jay Graham, has published over 300 colored photos of the architecture, furniture and handicrafts in Vietnam, Laos, and Cambodia, drawing insights into the combined influence of Indian, Chinese and French traditions in this region. More recently, Herbelin (2016) has taken an contrarian view by pointing out that the Indochine style was not entirely successful due to its high cost, which prompted the French to return to its classical design, such as in the case of the then Ministry of Finance (1927) and now Ministry of Foreign Affairs. Taking a departure from the popular praises of Saigon as "*la perle de l'Extrême Orient*" ("the pearl of the Far East") and Hanoi as comparable to Paris, Herbelin (2016) instead shows how the architecture in Vietnam under French colonialism was the result of negotiations and political strategies between various authorities, between colonial and local authorities, between the native population and the French, as well as between the different technical and aesthetic solutions that were offered at that time. The book reveals, thus, there is no such thing as a proper colonial architecture but just a phenomenon of hybridization, of intertwined cultures that contribute to the colonial moment (Thu Hang, 2017). Along this vein, Truong (2012) analyzes the harmonization of Eastern and Western elements in the Indochine style in Vietnam to conclude on there being a strong imprint of traditional Vietnamese architecture (*les traditions Vietnamiennes*). The Indochine style, the author argues, was after all born in Vietnam, and without doubt carries a large part of the native architectural elements, such as the use of wood and bamboo and the addition of balconies, verandahs, and internal corridors to accommodate the hot and humid tropical climate.

From a broader perspective, Hartingh, Craven-Smith-Milnes, and Tettoni (2007) review Vietnamese architecture from ancient to modern time and conclude that, although Vietnam's interior design has a touch of both Chinese and European cultures, the native design is in fact quite diverse thanks to the varying characteristics of different Vietnamese localities. The book showcases the diversity of Vietnamese culture, as evidenced by the myriad temples, shrines and pagodas across the country. Similarly, when looking at the traditions of Vietnamese architecture, Chu (2003) has documented a rich history of folkloristic architecture, from the way each household unit builds its house to the way a village and its folks contribute to the traditional architecture. On the evolving architecture in Vietnam, Herbelin (2016) has offered one potential explanation: besides the urbanization, the cityscape in the early twentieth century had changed dramatically because the French eradicated many laws on limiting household expenditure that were issued by the Nguyen dynasty earlier. As the people got to freely build two-storied houses and decorate them to their liking, these neoclassical designs were the results of influence from both France and China, where the East-West architectural fusion also took off strongly at the time (Herbelin, 2016).

For the purpose of this study, the team has summarized some notable ornamental characteristics in Table 1. These characteristics will form the basis of the analysis into the house façade in Hanoi.

**Table 1.** A summary of the non-Vietnamese and Vietnamese use of materials and ornamental designs

|  | **Non-Vietnamese** | **Vietnamese** |
| --- | --- | --- |
| **Materials** | Rock serves as one of the main materials in Western construction, such as in columns and balcony. | Rock is rarely used in house façade, instead it is used in sculpture of sacred animals. |
|  | Cement is also an important Western invention but is costlier and not as flexible as a construction material. | Ornamental designs are made primarily of ceramics. Meanwhile, houses are built of wood, and monuments are built of bricks. |
|  | The use of ceramics in ornamental design originated from southern China and became popular in Vietnam since the Nguyen dynasty (1802-1945). | The Vietnamese combination of honey, lime, paper pulp, and bagasse created a kind of dry mix mortar (*nề vữa*) suitable for construction. |
| **Ornaments** | Flowers of Chinese origins: lotus (Buddhism), chrysanthemum (Taoism), or peony (China's unofficial national follower) | In the façade of many houses in Hanoi, the pattern known as "Lily of the valley" (*Hoa huệ tháng Năm*) is the most popular. Examples:<br>- K Hospital today, earlier the Radium Institute of Indochina (1915-1920).<br>- Houses on 94 Hang Bong, 57 Hang Dieu, 161 Phung Hung, 23 |

| | Flowers and leaves associated with the West: lilies, tulip, olive branches, oak leaves, pine fruit, laurel branches, etc. | There is a dearth of Vietnamese research on the use of lily as an ornamental feature in architecture, although much has been written on the popularity of lily in the famous paintings of To Ngoc Van, Le Pho, Tran Van Can. | Nguyen Quang Bich, 144 Nguyen Thai Hoc (1920-1945) |
|---|---|---|---|
| | Seashell: an image seen from the ancient time, featured prominently in Western aesthetics. | The Vietnamese folk belief has long looked down on shells (*nghêu - sò - ốc - hến*), and thus, these images were never used in the native decorations. | |
| | While ancient Chinese people had used shells (贝) as a form of currency to exchange for precious things, only with some Western influence did China begin to use shells in ornamental designs. | With French influence, the image of seashell has appeared in some French architecture in Hanoi. Examples: <br> - The Workers' Theatre on Trang Tien street (earlier Cinema Palace). <br> - Cambodia Embassy (71a Tran Hung Dao). <br> - Buildings on 107 Tran Hung Dao, and 64-68 Ly Thuong Kiet, 190 Hang Bong. | |
| | Traditional Chinese symbols: wine gourd, coin, four sacred flowers, four sacred mythical creatures (Taoism), words such as 寿, 福, 乐, 喜, 万. | | |

*Hanoi Old Quarter and the house façade designs*

Upon its settlement in Hanoi in the late nineteenth century, the French soon began building its own streets and administrative offices. In 1883, the first governor issued a plan to turn some streets in the southeast of the Sword Lake (Ho Hoan Kiem)—equivalent with the streets of Le Phung Hieu, Ly Thai To, Le Thanh Ton, and Ngo Quyen today—fully into the style of French architecture. This area, dubbed by the Vietnamese as the "Western quarter," stands in contrast with the old commercial area that is known as Hanoi's 36 streets located in the north of the Sword Lake (Dinh & Groves, 2006). A recent review of French urban planning in Hanoi pointed out that the French had built 157 streets in their style here between 1884 and 1945, of which 74 were strictly French with houses built in accordance with European style (Phan, Nguyen, Dao, Ta, & Nguyen, 2017). Table 2 summarizes the three major periods of French influence over Hanoi architecture, based on the research by Tran (2011).

Table 2. An overview of the major historical events and characteristics of Hanoi architecture from 1860 to 1945, based on the research by Tran (2011).

| Time period | Historical notes | Hanoi architecture | Notable examples |
|---|---|---|---|
| **1860-1900** | Only French army engineers could design and build houses, thanks to their | Most constructions were for military residence, jail/prison, and church. | - Vietnam Military History Museum (earlier Hôtel du |

|  | experience in building army bases in Algeria. |  | quartier général de l'armée) (1877)<br>- Notre-Dame Cathedral Basilica of Saigon (1877-1880)<br>- St. Joseph's Cathedral in Hanoi (1883-1887) |
|---|---|---|---|
|  | French troops officially took over Hanoi in 1882. | "Early colonial architecture": all architectural designs were imported straight from France and Europe (Vietnam Associations of Architects, 2003). |  |
| **1900-1920** | The French began building administrative offices and houses in the city. | The dominant architecture is classical and neoclassical style with a touch of different French regions. | - Hanoi Presidential Palace (earlier, Residence of the Governor-General of French Indochina) (1901-1907)<br>- Vietnam People's Supreme Courthouse (earlier Courthouse) (1906).<br>- Hanoi Opera House (earlier Municipal Theatre) (1901-1911). |
|  | Rapid urbanization gave rise to many urban centers and resort towns from the north to the south. | Major administrative buildings were built in the grand style, all designs were strictly symmetrical. |  |
|  | The École Française D'Extrême-Orient (EFEO), French School of Asian Studies, was founded in 1900, headquartered in Hanoi. |  |  |
| **1920-1945** | This period saw an integration of East-West values in the local architecture, evidenced in the designs of the roof, console system, and various decorations. | The Indochine style (Ernest Herbrad, Charles Batteur): more attention paid to the roof, ventilation, patios, doors and windows; designs taking into account symbols originated from Buddhism, Confucianism, and Taoism. | - Vietnam Museum of History<br>- Vietnam Ministry of Foreign Affairs<br>- House No. 18 on Le Hong Phong Street |
|  | EFEO established a full Archeological Service, appointing both European and local staff as faculty and assistants in the preservation and restoration of historical monuments in Indochina (Clementin-Ojha & Manguin, 2007). | The Franco-Chinese style: more elaborate Chinese decorations in Western buildings, more emphasis on the chimney and the surface design of the roof. The windows and doors of this style do not have as many shutters as those in the Indochine style. |  |
|  |  | The Art Décor style: became popular in 1930-1945; known for its usage of geometric shapes, zigzag lines, bold color and patterns, and metallic/iron materials. | - State Bank of Vietnam Headquarters (earlier Banque de l'Indochine)<br>- Villa No. 9 on Le Hong Phong Street.<br>- Trang Tien Palaza (earlier Godard House) |

While Phan et al. (2017) noted that 80% of the inhabitants in the Western quarter were Vietnamese[1], it is important to keep in mind that since the first half of the twentieth century, there had always been a large number of foreign inhabitants, such as Chinese, Indian, and the biggest group of all – French (Tran, 2011). Studies on the house façade in Hanoi Old Quarter have taken mostly a qualitative approach, delving into the ornamental designs and their historical and aesthetic values, as shown by Tran (2011), Phan Cam Thuong, and Nguyen Duc Hoa. When looking at the house façade, one of the outstanding features is the combined use of Vietnamese national script (*Chữ Quốc Ngữ*), French, and Chinese (*Hán*) characters on advertisement billboards and entryways. The construction of many houses in Hanoi prior to 1945 shows the undeniable aesthetic appeal of having some Chinese texts on the façade. The display of written texts on the house façade, according to Tran (2011), was a practice originated from Europe. Thanks to such inscriptions, one could observe the transforming aesthetics in Hanoi, from pure European style to an integrated East-West style.

The extensive literature review here lays an important foundation for this study. The aesthetic features to be examined are not only highly representative of their originating cultures, whether that be French or Chinese, but are also reflective of the relationship between the architectural theories and practical application. On this basis, the research team could categorize the most notable variables and structure the coding of such data in its Bayesian model. The following section will go into details the materials and methods of this research.

**Materials and Methods**

*Materials*

This study started from 2007 when our artist, Bui Quang Khiem, started to take pictures of old houses in the Old Quarter of Hanoi. Each street in the 36 streets was originally home to a different trade, as reflected in the name of the street, such as Hang Bun for Rice Noodle Street or Hang Non for Conical Hat Street (Dinh & Groves, 2006). As the craftsmen and traders brought their village culture and customs to Hanoi, the city saw an emergence of buildings characteristic of traditional Vietnamese village life, such as the communal house (*đình*), village pagoda (*chùa*), or village gate (*cổng làng*). Most of the houses in these streets were built during the French colonial period and they exhibit the typical characteristics of the two-storied tube houses built around the twentieth century (Kien, 2008b). These houses are influenced by the architecture of the French houses and they also reflect the wealth of their owners (Dinh & Groves, 2006). From 2007 to 2018, Bui Quang Khiem has taken more than 500 pictures of the façades of the old houses around Hanoi. The process of photo-taking is meticulous and time-consuming, given the lively economic life of the city and the country (Vuong, 2014). Many of the houses are covered in advertisement boards and the photographer had to wait for the moment when the advertisements boards were put away. Many of the houses have been destroyed during the time and the pictures are the limited documents left of them. All of the pictures have been deposited openly online in the Open Science Framework (OSF)'s database and can be accessed at https://osf.io/tfy6k/.

---

[1] In 1948, there were 150,000 Hanoi inhabitants. The figure rose to 300,000 in 1951 and to 643,576 in 1960 (Nguyen, 2016).

**Figure 2:** Two examples of the photos of Hanoi's old houses used in this study

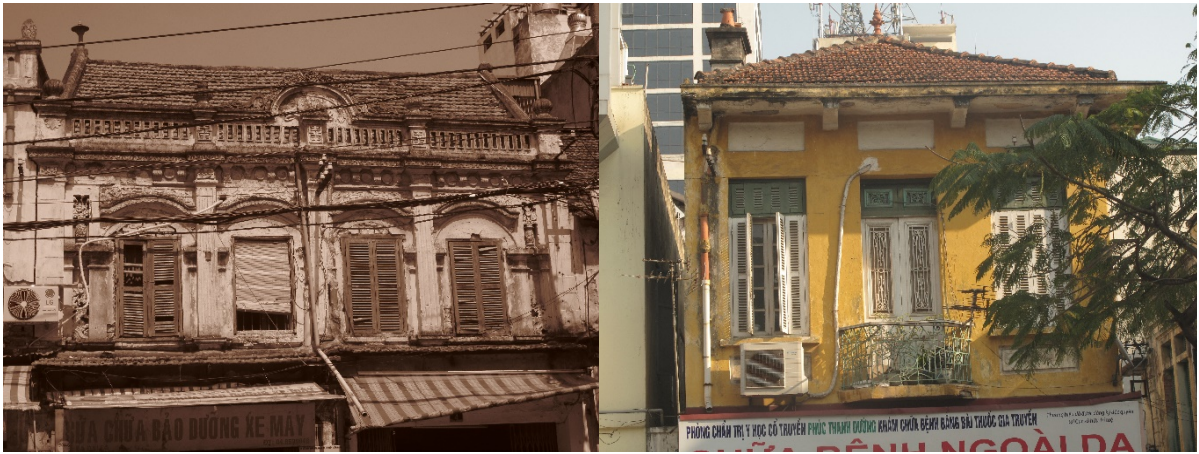

In September 2018, the research team started to select the pictures that are suitable for the purpose of this study. This study has left out the pictures that were not taken from a direct angle or did not show the decorative details or of not good quality. In the end, 248 photos of 68 old houses were chosen and encoded into 78 data lines in the excel sheet. The OSF's folder that store these photos is "Sorted images of Hanoi Houses", which can be accessed at https://osf.io/tfy6k/. The following section describes how the photos are encoded and the next section describe the statistical method employed in this study, the Bayesian Network analysis.

*Construction of variables*

The full dataset is stored in the file "FCCE1.181113.csv", which is deposited openly in OSF's sub-folder "Processing+Suppl" in the folder "Statistical Investigations" [Doi: https://osf.io/tfy6k/].

Dependent variables:

- **TR: Short for "The traditional feeling."** This categorical variable takes on one of these values strong, medium or none (coded as *"TR_S," "TR_M"* and *"TR_N"* respectively)

- **MD: Short for "The modern feeling."** This categorical variable takes on one of these values strong, medium or none (coded as *"MD_S," "MD_M"* and *"MD_N"* respectively). Strong means the façade brings a feeling of French-ness, Medium means the façade is a good mixture of Franco and Chinese style, none means the French-ness is unclear or none at all.

- **CE: Short for "Cultural evolution."** This categorical variable is about the cultural evolution and acculturation being represented on a façade. It takes on one of these values: beginning, strong representation of cultural evolution (evolving), or complete (coded as *"C_B," "C_E"* and *"C_C"* respectively). When a façade looks traditional, it is judged the cultural evolution and acculturation process just started *("C_B")*. When a façade looks like there is a tradition from the traditional and the modern, it is judged the cultural evolution and acculturation process is evolving *("C_E")*. When a façade looks like it is completely modern and similar to the houses in modern Hanoi, it is judged the cultural evolution and acculturation process has been completed *("C_C")*. All of the team members participate in making this judgement, each makes their decisions in choosing the value for the houses independently.

In the initial model, the cultural evolution variable is probabilistically dependent on both the traditional and the modern feeling.

Independent variables:

- **B: Short for "Buddhism-inspired decorations."** This categorical variable takes on one of these values: strong, none or weak (coded as *"B_S," "B_N"* and *"B_W"* respectively). Based on the appearances of Buddhism-inspired patterns/symbols such as flower, peach, lotus, wheel, etc., we rate how strong the characteristic of Buddhism on the façade is. When patterns/symbols are in the center, big or repeatedly used, it is rated as strong.

- **T: Short for "Taoism-inspired decorations."** This categorical variable takes on one of these values: strong, none or weak (coded as *"T_S," "T_N"* and *"T_W"* respectively). Based on the appearances of Taoism -inspired patterns/symbols such as cloud, Ba Gua mirror, yin-yang, etc., we rate how strong the characteristic of Taoism on the façade is. When patterns/symbols are in the center, big or repeatedly used, it is rated as strong.

- **C: Short for "Confucianism-inspired decorations."** This categorical variable takes on one of these values: strong, none or weak (coded as *"C_S," "C_N"* and *"C_W"* respectively). Based on the appearances of Confucianism -inspired patterns/symbols such as dragon, Chinese characters, paper rolls, etc., we rate how strong the characteristic of Confucianism on the façade is. When patterns/symbols are in the center, big or repeatedly used, it is rated as strong.

- **DC: Short for "Decoration."** This categorical variable takes on one of these values: French, Chinese or Hybrid (coded as *"DC_FR," "DC_CN"* and *"DC_HY"* respectively).

In the initial model, the decoration variable (DC) is probabilistically dependent on the C, B, and T variables.

- **DO: Short for "Door."** This categorical variable takes on one of these values: French, Chinese or Hybrid (coded as *"DO_FR," "DO_CN"* and *"DO_HY"* respectively).

- **PL: Short for "Pillar."** This categorical variable takes on one of these values: French, Chinese or Hybrid (coded as *"PL_FR," "PL_CN"* and *"PL_HY"* respectively).

- **RF: Short for "Roof."** This categorical variable takes on one of these values: French, Chinese or Hybrid (coded as *"RF_FR," "RF_CN"* and *"RF_HY"* respectively).

In the initial model, the door, pillar and roof (DO, PL and RF) variables probabilistically influence the modern feeling (MD) variable. The decoration variable (DC) also probabilistically influences both the traditional (TR) and the modern (MD) feeling variable.

*The initial conceptual model*

Figure 3 presents the initial Bayesian network model for the probabilistic dependency among the variables. DC is probabilistically dependent on C, T and B. MD is probabilistically dependent on DC, RF, PL, and DO. TR is probabilistically dependent on DC. CE is probabilistically dependent on TR and MD. These relationships are encoded into a directed acyclic graph (DAG) and its visualization can be seen in Figure 3.

**Figure 3:** The initial Bayesian network model. A directed acyclic graph which represents the probabilistic dependency among the variables.

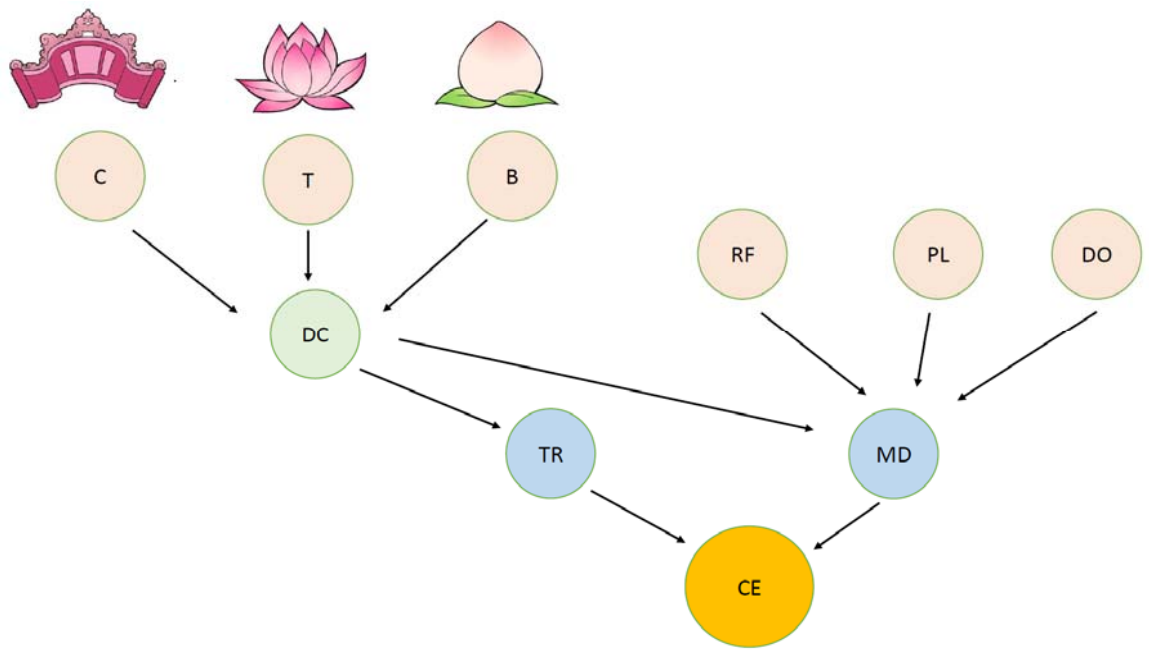

This study deployed the **bnlearn** package (short for "Bayesian network learning") to create the DAG, for further technical explanation, see *Bayesian Networks- With Examples in R* of Scutari and Denis (2014).

To create the DAG, the following codes are run in the program R (v.3.3.1).

```
library(bnlearn)

dag <- empty.graph(nodes = c("B", "T", "C","DC","RF","DO","PL","MD","TR","CE"))
nodes(dag)
```

It is important to note that in the beginning the DAG is an empty graph, meaning there is no probabilistic dependencies encoded in it. The next step is to add the relationships among the variables.

```
dag <- set.arc(dag, from = "B", to = "DC")
dag <- set.arc(dag, from = "T", to = "DC")
dag <- set.arc(dag, from = "C", to = "DC")

dag <- set.arc(dag, from = "RF", to = "MD")
dag <- set.arc(dag, from = "PL", to = "MD")
dag <- set.arc(dag, from = "DO", to = "MD")

dag <- set.arc(dag, from = "DC", to = "TR")
dag <- set.arc(dag, from = "DC", to = "MD")

dag <- set.arc(dag, from = "TR", to = "CE")
dag <- set.arc(dag, from = "MOD", to = "CE")
```

To test whether the DAG is correct, the plot function can be used. The result can be seen in figure 4.

```
modelstring(dag)
[1] "[B][T][C][RF][DO][PL][DC|B:T:C][MD|DC:RF:DO:PL][TR|DC][CE|MOD:TR]"

plot(dag)
```

**Figure 4:** The DAG created using the bnlearn package. This directed acyclic graph is identical to the one presented in Figure 1.

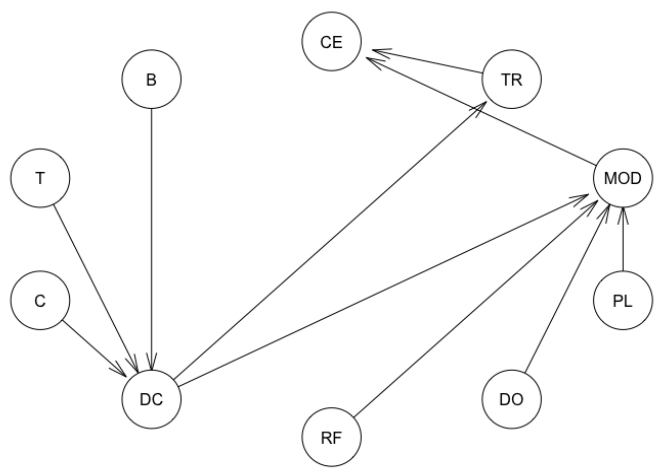

The formal definition of how the dependencies are encoded in the map into the probability space via conditional independence relationships is provided in Equation 1:

$$\Pr(C, T, B, DC, RF, PL, DO, MD, TR, CE) =$$
$$\Pr(C)\Pr(T)\Pr(B)\Pr(RF)\Pr(PL)\Pr(DO)\Pr(DC|C,T,B)\Pr(MD|DC,RF,PL,DO)\Pr(TR|DC)\Pr(CE|MD,TR)$$

The conditional probabilities in the local distributions can be estimated with the empirical frequencies in the datasets, for example, Equation 2 shows the conditional probabilities of a façade exhibits a strong presence of cultural evolution given it also gives off a strong feeling of modern:

$$\widehat{PR}(CE = CE_E | MD = MD\_S) = \frac{\widehat{Pr}(CE = \text{"CE\_E"}, MD = MD\_S)}{\widehat{Pr}(MD = MD\_S)}$$
$$= \frac{\text{number of observations for which } CE = CE\_E \text{ and } MD = MD\_S}{\text{number of observations for which } MD = MD\_S}$$

This is the classic frequentist and maximum likelihood estimates, for further discussions, see (Scutari & Denis, 2014). In this study, in order to evaluate the structure of the Bayesian networks, the Bayesian Information Criterion (BIC) score (Equation 3) is employed:

$$BIC = \log \widehat{Pr}(C, T, B, DC, RF, PL, DO, MD, TR, CE) - \frac{d}{2}\log n =$$
$$= \left[\log \widehat{Pr}(C) - \frac{d_C}{2}\log n\right] + \left[\log \widehat{Pr}(B) - \frac{d_B}{2}\log n\right] + \left[\log \widehat{Pr}(T) - \frac{d_T}{2}\log n\right] +$$
$$\left[\log \widehat{Pr}(RF) - \frac{d_{RF}}{2}\log n\right] + \left[\log \widehat{Pr}(PL) - \frac{d_{PL}}{2}\log n\right] + \left[\log \widehat{Pr}(DO) - \frac{d_{DO}}{2}\log n\right] +$$
$$\left[\log \widehat{Pr}(DC|C,B,T) - \frac{d_{DC}}{2}\log n\right] + \left[\log \widehat{Pr}(MD|DC,RF,PL,DO) - \frac{d_{MD}}{2}\log n\right] +$$
$$\left[\log \widehat{Pr}(TR|DC) - \frac{d_{TD}}{2}\log n\right] + \left[\log \widehat{Pr}(TR|DC) - \frac{d_{TD}}{2}\log n\right] +$$
$$\left[\log \widehat{Pr}(CE|MD,TR) - \frac{d_{MD}}{2}\log n\right]$$

# Results

*Preliminary evaluation: a search for possible and meaningful causal models*

First, the strength of probabilistic dependence corresponding to each arc in the initial model is tested using the ***arc.strength*** function in **bnlearn** package. The test returns non-significant results ($p>0.05$) across the network. The following box details the codes and the results.

```
strength = arc.strength(dag, data = data1, criterion = "x2")

  from to     strength
1    B DC   9.763066e-01
2    T DC   1.000000e+00
3    C DC   9.999153e-01
4   RF MD   1.000000e+00
5   PL MD   1.000000e+00
6   DO MD   1.000000e+00
7   DC TR   3.093218e-08
8   DC MD   1.000000e+00
9   TR CE   7.989842e-01
10  MD CE   9.174464e-01
```

With the initial Bayesian network model, whose Bayesian Information criterion (BIC) score equals -979.0649 and Bayesian Dirichlet equivalent uniform (BDen) score equals -552.611, not showing any significance, 200 random DAGs are generated using machine algorithms for a further investigation of the relationship among the variables. Table 3 provides 21 DAGs that satisfy the technical requirements. The initial criteria for generating the DAGs are: (i) there are connections to CE; (ii) there is no connection from CE; and (iii) the minimum number of connections is at least five.

**Table 3:** Twenty-one DAGs with the best BIC scores are presented. Note that all of the scores are better than the initial model.

| (M1) -527.5334019 [MD][PL][T][TR|MD][B|TR][RF|TR][C|RF][DC|B][DO|RF][CE|DO] | (M2) -525.1395526 [B][DO][PL][T][DC|B][CE|DC][RF|DC][C|RF][TR|C][MD|TR] | (M3) -528.9434423 [C][PL][RF][DO|RF][MD|RF][TR|RF][CE|DO][DC|TR][B|DC][T|B] |
|---|---|---|

(M4) -519.0916518
[B][MD][PL][T][DC|B][RF|DC][CE|RF][DO|RF][TR|RF][C|TR]

(M5) -521.5010174
[MD][PL][T][TR|MD][B|TR][C|TR][DC|TR][CE|DC][RF|DC][DO|RF]

(M6) -521.5731987
[MD][PL][TR|MD][DC|TR][B|DC][C|DC][CE|DC][RF|DC][DO|RF][T|B]

(M7) -517.3542449
[B][DO][PL][T][DC|B][C|DC][CE|DC][RF|DC][TR|DC][MD|TR]

(M8) -519.8741391
[B][T][DC|B][C|DC][MD|DC][RF|DC][TR|DC][CE|RF][DO|RF][PL|MD]

(M9) -523.3763526
[B][PL][DC|B][T|B][C|DC][CE|DC][MD|DC][RF|DC][TR|DC][DO|RF]

(M10) -522.0605958
[B][PL][T][C|T][DC|B][RF|DC][CE|RF][DO|RF][TR|RF][MD|TR]

(M11) -524.8840612
[C][DO][MD][PL][T][TR|C][B|TR][DC|TR][CE|DC][RF|DC]

(M12) -526.1012981
[B][MD][PL][T][TR|MD][C|TR][DC|TR][DO|TR][CE|DC][RF|DC]

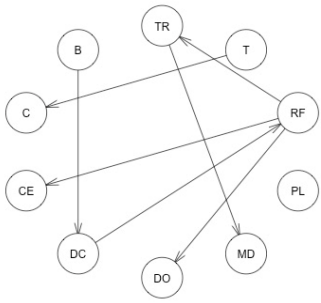
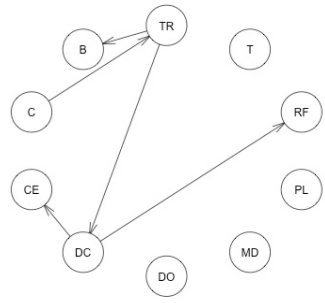
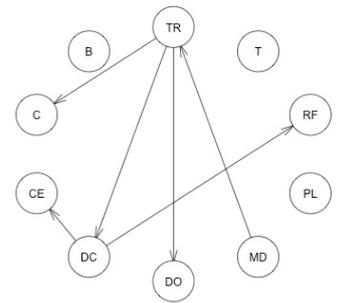

(M13) -517.126813
[B][C][DO][PL][T][DC|B][CE|DC][RF|DC][TR|RF][MD|TR]

(M14) -531.6963744
[PL][T][C|T][TR|C][B|TR][RF|TR][CE|RF][DC|B][DO|RF][MD|DC]

(M15) -527.3729879
[B][DC|B][C|DC][MD|DC][RF|DC][TR|DC][CE|RF][DO|RF][T|C][PL|DO]

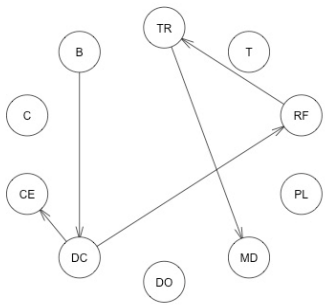
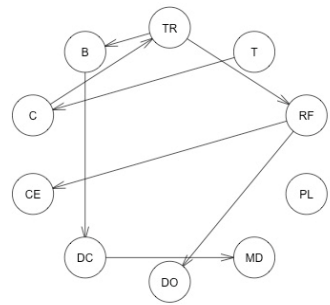
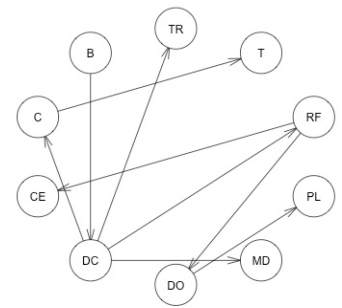

(M16) -530.1548123
[B][DC|B][T|B][PL|DC][RF|DC][DO|RF][TR|RF][C|TR][MD|TR][CE|MD]

(M17) -530.3505663
[DC][CE|DC][MD|DC][RF|DC][TR|DC][B|TR][C|TR][DO|RF][PL|MD][T|B]

(M18) -527.3428835
[B][DC|B][C|DC][TR|DC][MD|TR][RF|TR][T|C][CE|RF][DO|RF][PL|MD]

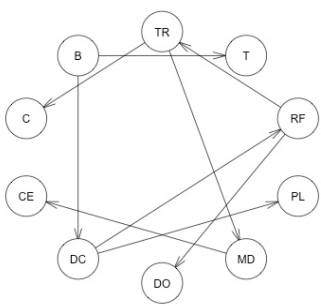
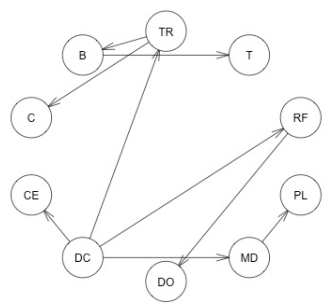
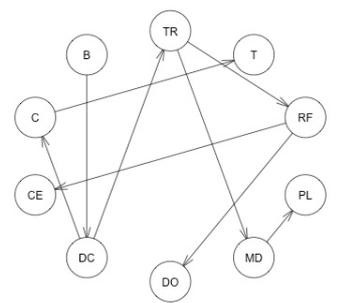

(M19) -528.3939493
[PL][MD|PL][TR|MD][B|TR][DC|B][T|B][C|DC][CE|DC][RF|DC][DO|RF]

(M20) -529.9446029
[MD][TR|MD][B|TR][C|TR][DC|B][PL|B][T|B][RF|DC][CE|RF][DO|RF]

(M21) -531.4081897
[B][T][C|T][DC|B][MD|DC][RF|DC][CE|RF][DO|MD][PL|MD][TR|MD]

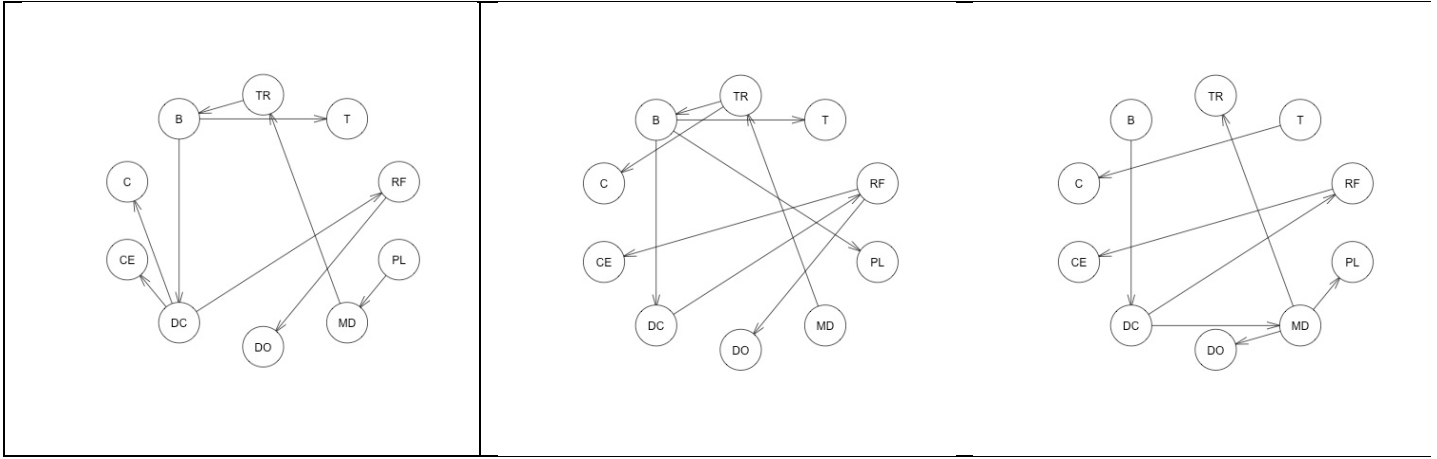

The following box shows the codes to compute the scores for the generated networks and to select the ones with the best BIC scores.

```
# computing scores from the generated networks
scoreNetwork <- function(nets,data,type = "bic") {
        scNet = c()
        stNet = c()
        for(n in 1:length(nets))
        {
                sc = score(nets[[n]], data = data, type = type)
                scNet <- c(scNet,sc)
                stNet <- c(stNet,modelstring(nets[[n]]))
        }
                return(list('score'=scNet,'model'=stNet))
}
# selecting top 5 models with highest scores among 100
topNetwork <- function(scores,top) {
        sc <- tail(order(scores$score),top)
        scNet = scores$score[sc]
        stNet = scores$model[sc]
                return(list('score'=scNet,'model'=stNet))
}
scores <- scoreNetwork(genNets,data1)
topNet <- topNetwork(scores,5)
```

The top five models were chosen on the basis that they have higher BIC network scores and they are M13, M7, M4, M8, and M5.

*Inference from top four models*

It is notable that all models start with the dependency of decoration (DC) on Buddhism-inspired decorative patterns (B). In other words, the value DC takes (Chinese, French or Hybrid) is probabilistically dependent on the value of B (none, weak or strong). Secondly, in all the models, there are some form of interactions between the decoration (DC) variable and Cultural evolution (CE) variable. For example, in Table 4, for M13 and M7, CE is directly dependent on DC; in M4 and M8, CE is indirectly dependent on DC through the roof design (RF). All four models are statistically significant under the frequentist approach, however, as they can be categorized in the two patterns above, the next section will focus on the best model for each pattern: M4 and M13.

Table 4: Four models with the best network scores are presented.

| Score | Arc Strength (x2) | Arc Strength (mi) | ci.test (mi) | ci.test (x2) |
|---|---|---|---|---|
| (M13) -517.13 | B->DC 4.51E-06<br>DC->CE 2.98E-06<br>DC->RF 1.35E-13<br>RF->TR 8.59E-07<br>TR->MD 1.66E-06 | B->DC 8.86E-06<br>DC->CE 6.83E-04<br>DC->RF 1.01E-06<br>RF->TR 1.55E-05<br>TR->MD 1.41E-04 | ci.test("CE", "DC",c("B"), test = "mi", data = data1)<br><br>data: CE ~ DC \| B<br>mi = 21.477, df = 12, p-value = 0.04381 | ci.test("CE", "DC",c("B"), test = "x2", data = data1)<br><br>data: CE ~ DC \| B<br>x2 = 19.145, df = 12, p-value = 0.08509 |
| (M7) -517.35 | B->DC 4.51E-06<br>DC->C 8.69E-03<br>DC->CE 2.98E-06<br>DC->RF 1.35E-13<br>DC->TR 3.09E-08<br>TR->MD 1.66E-06 | B->DC 8.86E-06<br>DC->C 3.01E-03<br>DC->CE 6.83E-04<br>DC->RF 1.01E-06<br>DC->TR 1.01E-05<br>TR->MD 1.41E-04 | ci.test("CE", "DC",c("B"), test = "mi", data = data1)<br><br>data: CE ~ DC \| B<br>mi = 21.477, df = 12, p-value = 0.04381 | ci.test("CE", "DC",c("B"), test = "x2", data = data1)<br><br>data: CE ~ DC \| B<br>x2 = 19.145, df = 12, p-value = 0.08509 |
| (M4) -519.09 | B->DC 4.51E-06<br>DC->RF 1.35E-13<br>RF->CE 3.37E-07<br>RF->DO 3.92E-09<br>RF->TR 8.59E-07<br>TR->C 2.17E-03 | B->DC 8.86E-06<br>DC->RF 1.01E-06<br>RF->CE 3.50E-04<br>RF->DO 8.80E-04<br>RF->TR 1.55E-05<br>TR->C 3.03E-03 | ci.test("CE", "RF",c("DC"), test = "mi", data = data1)<br><br>data: CE ~ RF \| DC<br>mi = 15.966, df = 12, p-value = 0.1928 | ci.test("CE", "RF",c("DC"), test = "x2", data = data1)<br><br>data: CE ~ RF \| DC<br>x2 = 25.327, df = 12, p-value = 0.01335 |
| (M8) -519.87 | B->DC 4.51E-06<br>DC->C 8.69E-03<br>DC->MD 2.16E-09<br>DC->RF 1.35E-13<br>DC->TR 3.09E-08<br>RF->CE 3.37E-07<br>RF->DO 3.92E-09<br>MD->PL 2.37E-16 | B->DC 8.86E-06<br>DC->C 8.69E-03<br>DC->MD 7.31E-04<br>DC->RF 1.01E-06<br>DC->TR 1.01E-05<br>RF->CE 3.50E-04<br>RF->DO 8.80E-04<br>MD->PL 1.08E-02 | ci.test("CE", "RF",c("DC"), test = "mi", data = data1)<br><br>data: CE ~ RF \| DC<br>mi = 15.966, df = 12, p-value = 0.1928 | ci.test("CE", "RF",c("DC"), test = "x2", data = data1)<br><br>data: CE ~ RF \| DC<br>x2 = 25.327, df = 12, p-value = 0.01335 |

In model M13, when Buddhism-inspired decorative patterns are strong (B_S), the probability of the Decoration variable (DC) takes on the value of hybrid (HY) and (CN) increases compared with when B equals none (B_N). When there is a weak presence of Buddhism-inspired decorative patterns/symbols, DC most likely takes on the value hybrid (HY) (90%). For the variable Cultural evolution (CE), when decoration of a façade is hybrid, the probability of the cultural evolution process is happening (CE=CE_E) is highest (90%). When the decoration is judged to be Chinese, the probability of CE just started equals roughly 60%. When the decoration is judged to be French, the probability of CE process has finished equals around 55%.

Figure 5: Probability distribution of the categorical variables in model M13 given different evidences.

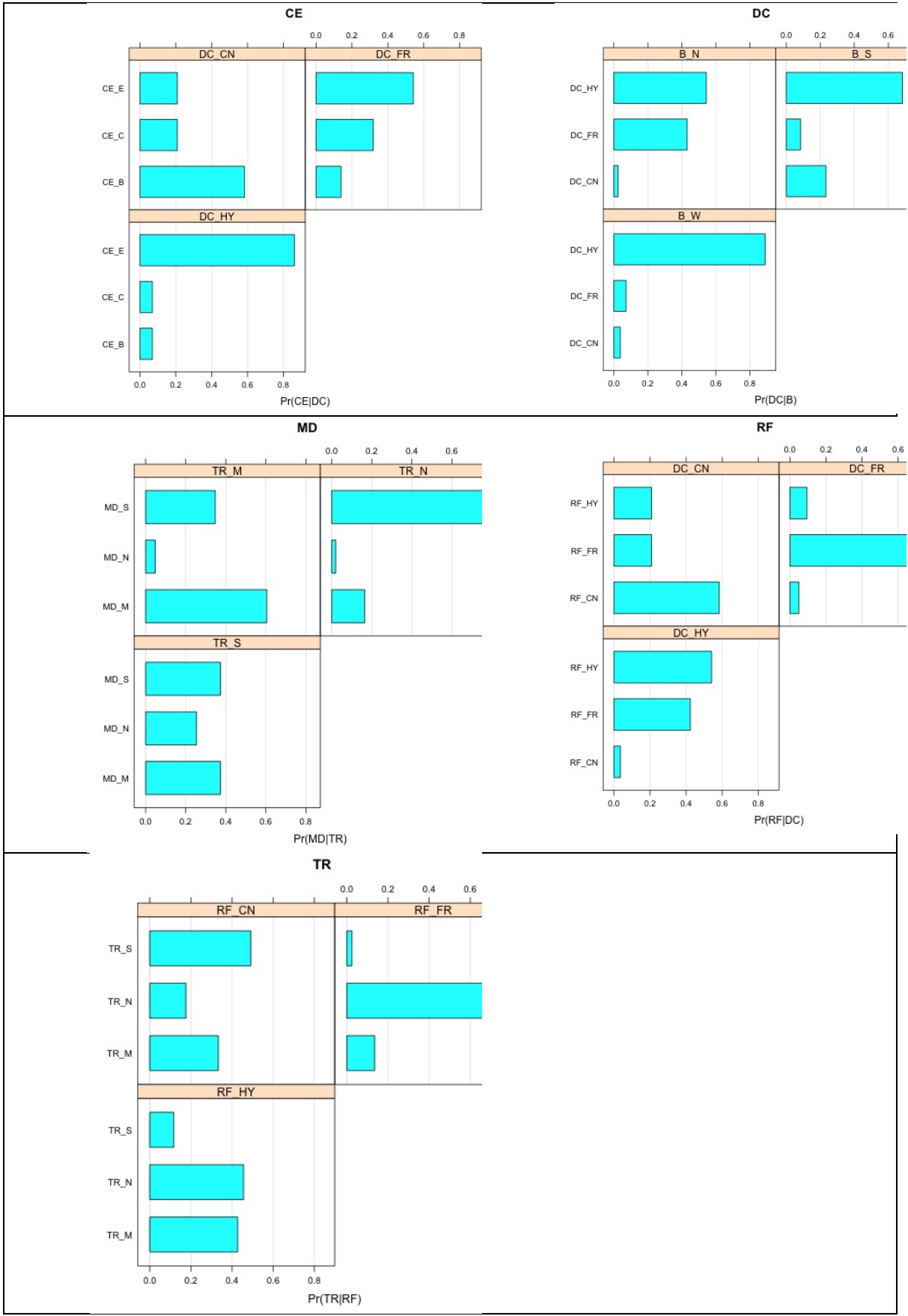

In Figure 6, the judgement on the cultural evolution process (CE) is dependent on whether the roof (RF) is Chinese, French or Hybrid. When the roof is Chinese style, the probability of CE process just started is the highest, 60%. When the roof is hybrid, the probability of CE process is happening is more than 80%.

**Figure 6:** Probability distribution of the categorical variables in model M4 given different evidences.

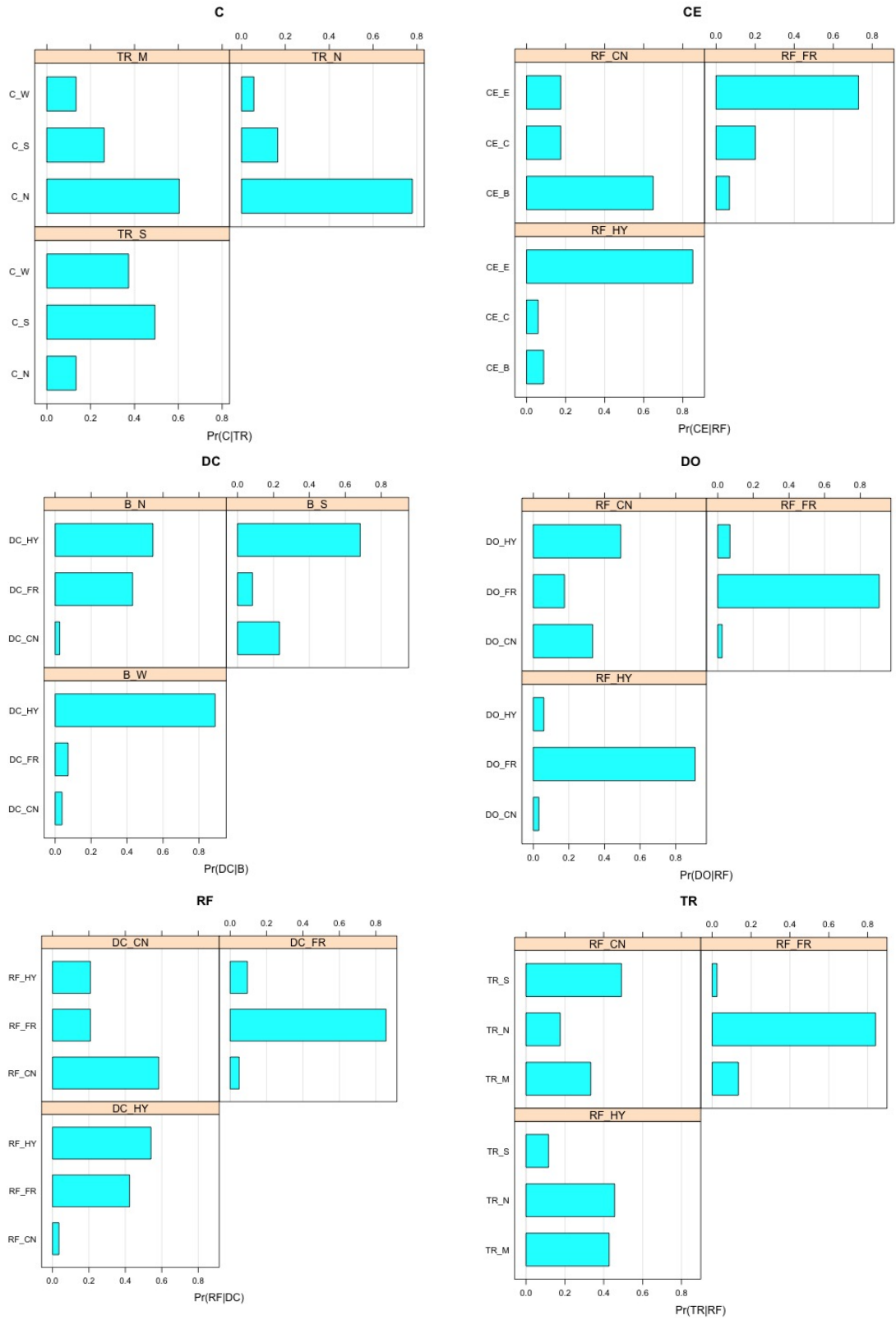

*Robustness verification with Hamiltonian MCMC pondering for Bayesian networks: JAGS and Stan*

In this section, the four best models are verified using the Hamiltonian MCMC method. One of the strengths of the Bayesian statistics approach is that it can generate technical figures for verification (Kruschke, 2015; McElreath, 2016). Using both the JAGs and Stan, this study diagnoses the robustness of the M4 and M13 Bayesian networks. The estimation has in total 20,000 iterations divided into 4 Markov chains. The file that contains all of the results (including the results for model M8, M7 and M5) is deposited in OSF's "Statistical Investigations" folder, "Files" sub-folder [URL: https://osf.io/tfy6k/].

**Figure 7:** M4's coefficients' posterior distribution

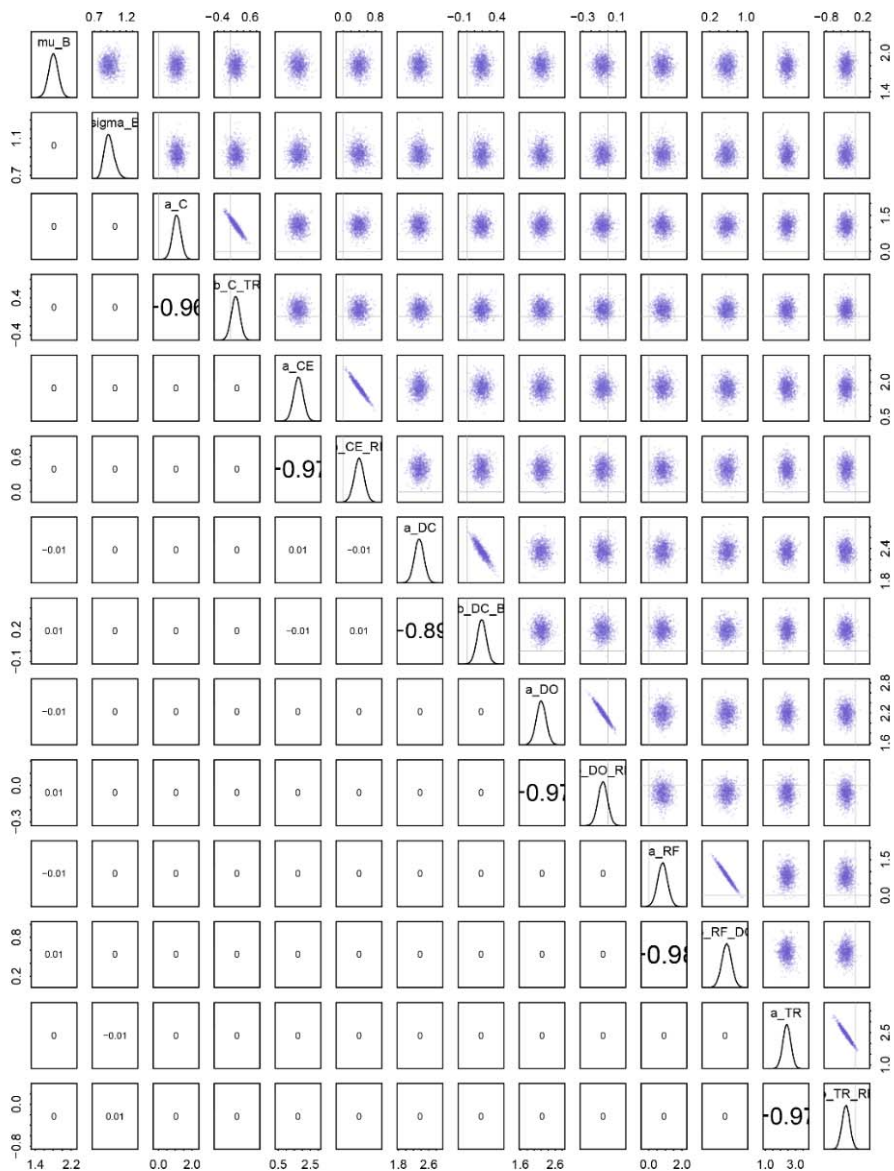

The posterior distributions of all coefficients for model M4, as shown in Figure 7, all satisfy the standard distribution. In Figure 8, we presented an example of testing the validity of the coefficients $\beta_{\{CE \cdot RF\}}$ in M4, where CE is dependent on RF probabilistically. Here, one can see the chains fluctuate around 0.4 and has a good-mix (Figure 8, top left). For the autocorrelation function, the four chains converge very quickly after lag 3 and the effective sample size (ESS) is nearly 66,700, indicating computational efficiency. Shrinking factor of computed mean values converged to 1.0 quite fast, while the Monte Carlo standard error (MCSE) is less than 0.05%.

**Figure 8:** Hamiltonian MCMC technical validations for $\beta_{\{CE \cdot RF\}}$ in model 4 using Stan codes.

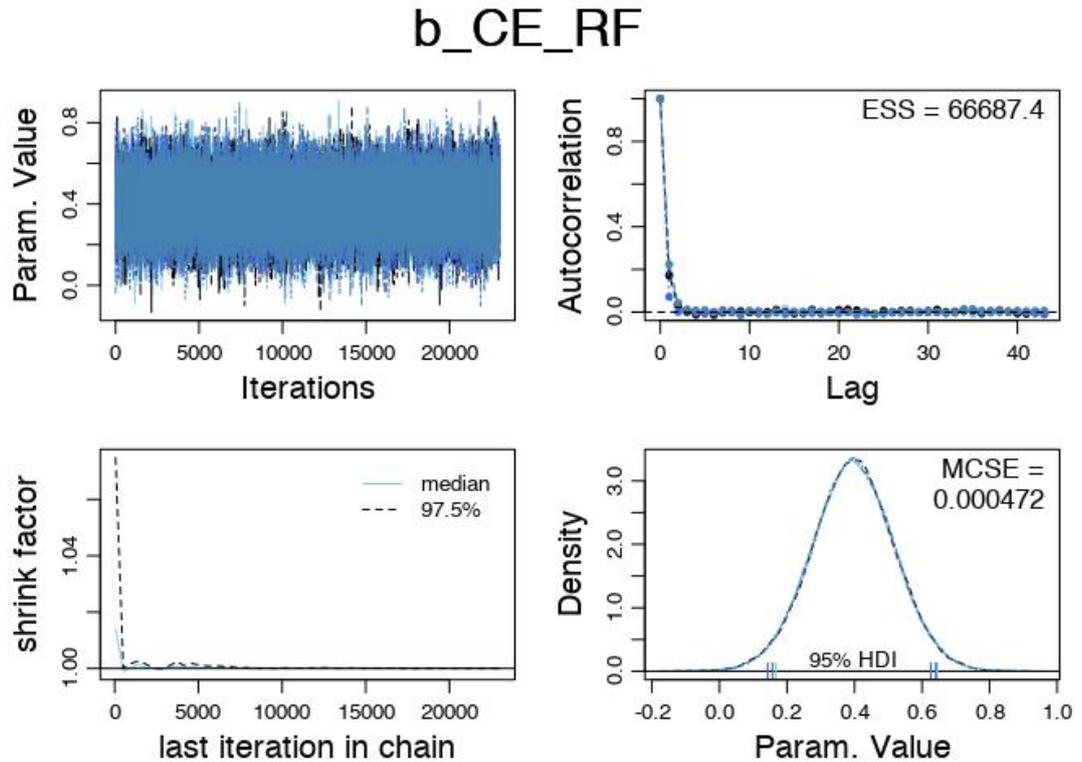

Similar to M4, the model M13 is also verified using the MCMC method. Figure 7 shows all posterior distributions of the coefficients of this model, all satisfy the technical standard. Similar to M4, the model M13 is also verified using the MCMC method. Figure 9 shows all posterior distributions of the coefficients of this model, all of them satisfy the technical requirement. Figure 10 shows the Hamiltonian MCMC diagnostics for the two coefficients of the cultural evolution variables in M13. All of the technical measurements indicate convergence of the posteriors.

**Figure 9:** M4's coefficients' posterior distribution

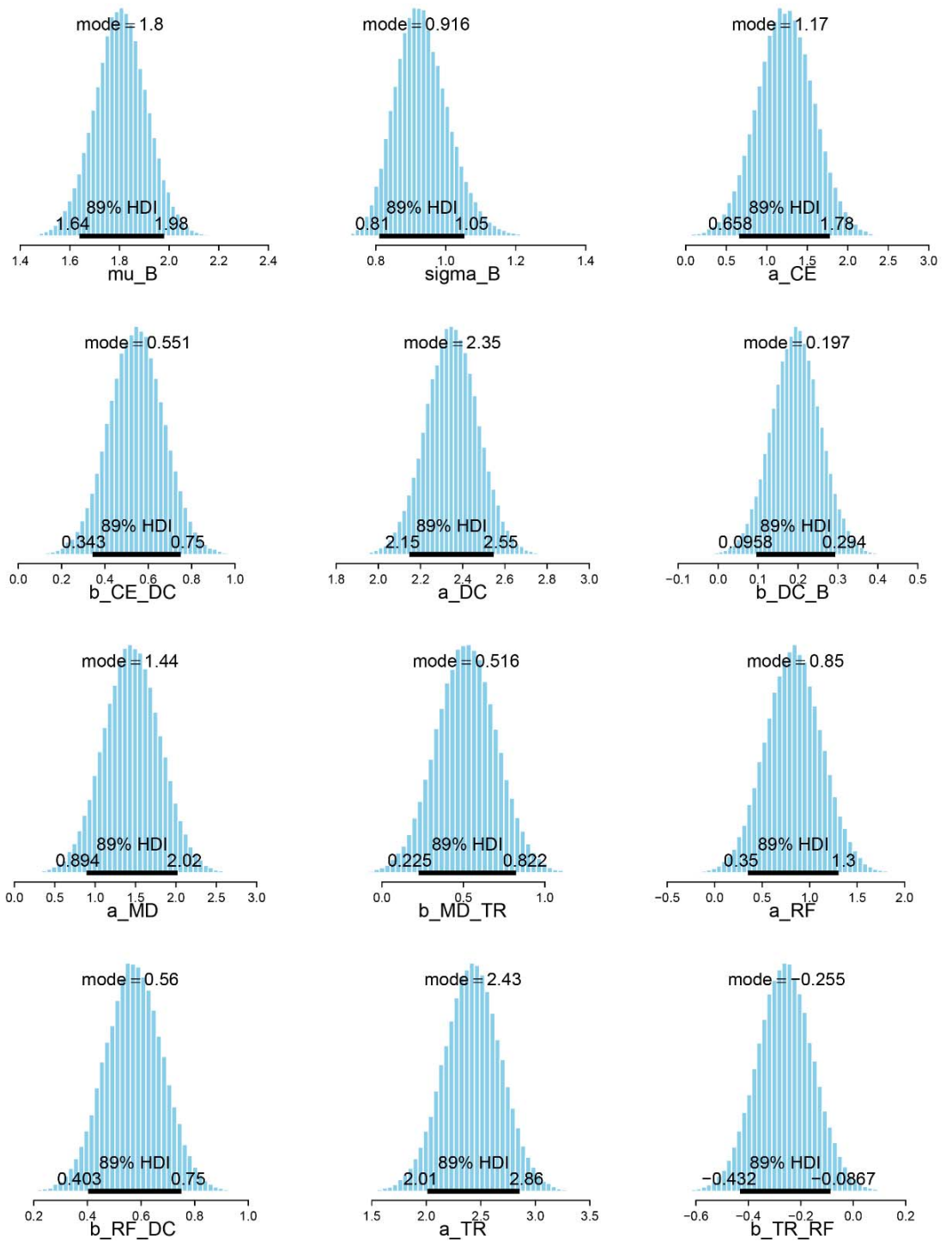

**Figure 10:** Hamiltonian MCMC technical validations for $\beta_{\{CE\}}$ in model 4 using JAGS codes

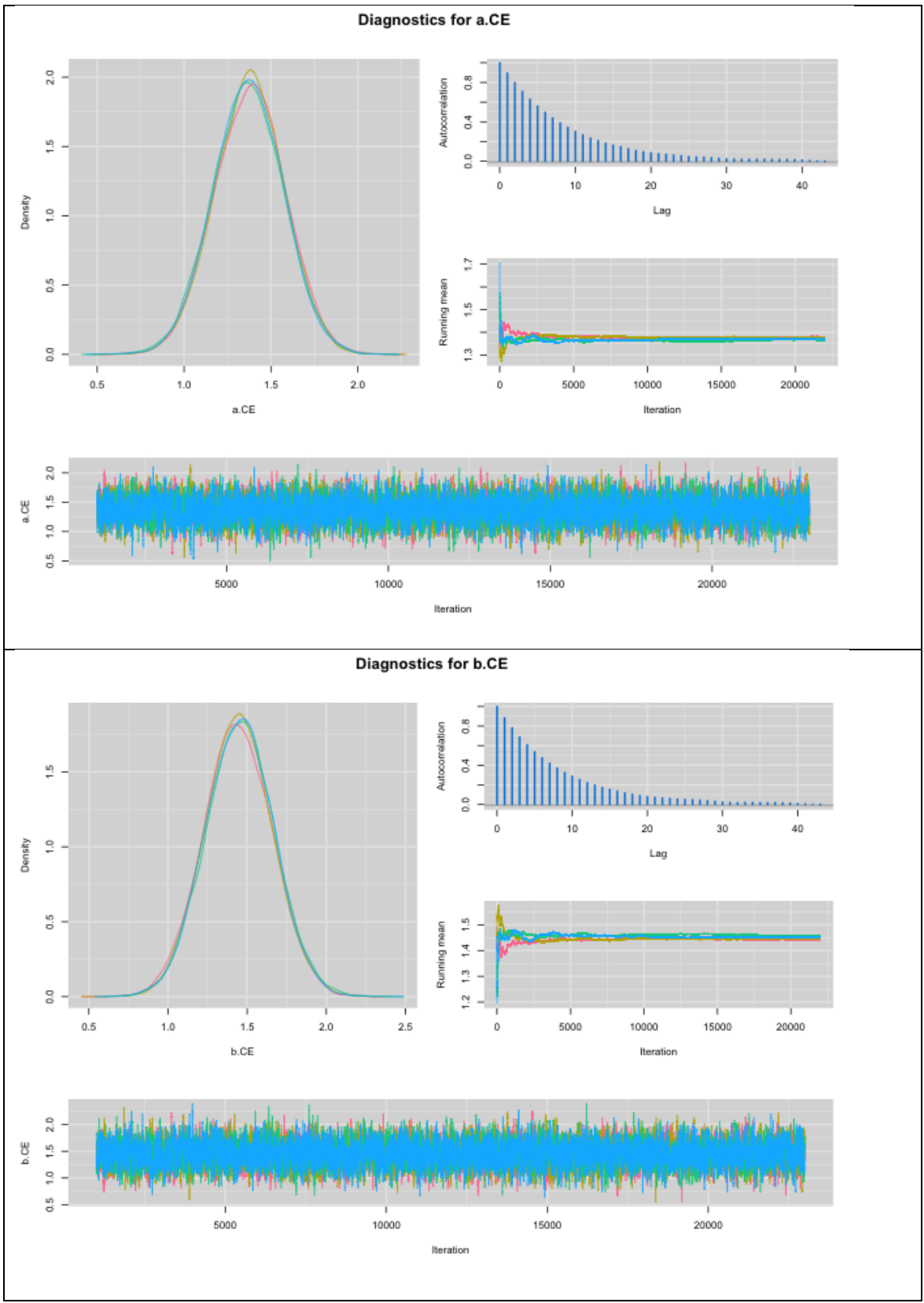

## Discussion

The findings in this study contribute to the literature on Franco-Chinese influence in Southeast Asian architecture in general and in Vietnam in particular, especially in its diverging methodology from the mainly qualitative approach (Dinh & Groves, 2006;

Hartingh et al., 2007; Herbelin, 2016; Le, 2013; Nguyen, 2014; Nguyen, 2016; Phan et al., 2017; Tran, 2011; Tran & Nguyen, 2012; Truong, 2012; Vietnam Associations of Architects, 2003; Vongvilay et al., 2015; Walker, 2011). Through the Bayesian networks analysis, the study shows that, despite a small volume of data and a coding of highly representative variables, it is nonetheless plausible to find the impacts of certain cultural elements on the aesthetics, architecture decisions of people at that time. By giving equal value to every input in judging the Franco-Chinese or hybrid modern feeling of a house façade, the study takes the mean or mode in Bayesian posterior distribution, ruling out the possibility of any "expert opinion" being more valuable. This approach is in line with the probabilistic interpretation of regularization in Bayesian statistics, and thus, shows its rich potential in social sciences studies as a whole (Kruschke, 2015; McElreath, 2016; Vuong, La, et al., 2018).

It is important to note here that, although the findings do not confirm the presumed model (Figure 1), they do answer to the two research questions that this study starts out with. For research question No.1, the in-depth technical analysis of this study shows the plausibility of judging the aesthetics, architecture and designs of the house façade in Hanoi to find traces of cultural evolution in the early 20$^{th}$ century in Vietnam. By highlighting ornamental features that are highly representative of French and Chinese cultures as well as their hybridity, the study was able to construct an efficient Bayesian model that draws out the association and correlation among different variables.

For research question No.2 on which elements most affect the Vietnamese perception on cultural evolution of Hanoi architecture in early 20$^{th}$ century, the Bayesian networks investigation into Hanoi architecture indicates a strong influence of Buddhism over the decorations of the house façade. In the top 5 networks with the best BIC scores and small *p*-values, the variable DC always has a direct probabilistic dependency on the variable B. Given the predominance of Confucianism in Vietnamese culture (Vuong, 2016; Vuong & Tran, 2009), this result is quite interesting. Two possible explanations could account for this finding. First, the lack of Confucian presence is attributable to the end of *Hán* script in academic and official settings in 1919 and the elevation to national status of *Chữ Quốc Ngữ* in Vietnam in 1945 (Chiung, 2001; Trinh, 2000). Second, it is possible that some of the French architects who were involved in the planning and building of Hanoi at the time had been influenced by Buddhism during their work with the École Française d'Extrême-Orient (EFEO), the French School of Asian Studies founded in 1900. As Clementin-Ojha and Manguin (2007) noted, in 1920, the EFEO established a full Archeological Service and Charles Batteur was tasked with the restoration of Vietnamese antiquities. In 1922, Charles Batteur helped restore the Pagoda of the Single Pillar (*Chùa Một Cột*) in Hanoi and numerous other pagodas that were damaged by a typhoon in 1929 (Clementin-Ojha & Manguin, 2007). The architects' involvement in preserving and restoring the Buddhist pagodas may help explain the somewhat prominent feature of Buddhist decorations in the house façade in Hanoi.

**Conclusion**

The inquiry into the house façade of Hanoi's Old Quarter has brought sharp focus onto the ornamental features and their cultures of influence. The topic itself, as this study shows, has been the subject of numerous qualitative research, yet none has applied

Bayesian inference to systematically and rigorously document the selection of such ornamental designs and the perceived source of cultural influence. This study, therefore, opens up a new approach for social sciences in general and for study of cultural evolution and architectural transformation in particular.

**Figure 11:** A linocut painting by courtesy of Bui Quang Khiem

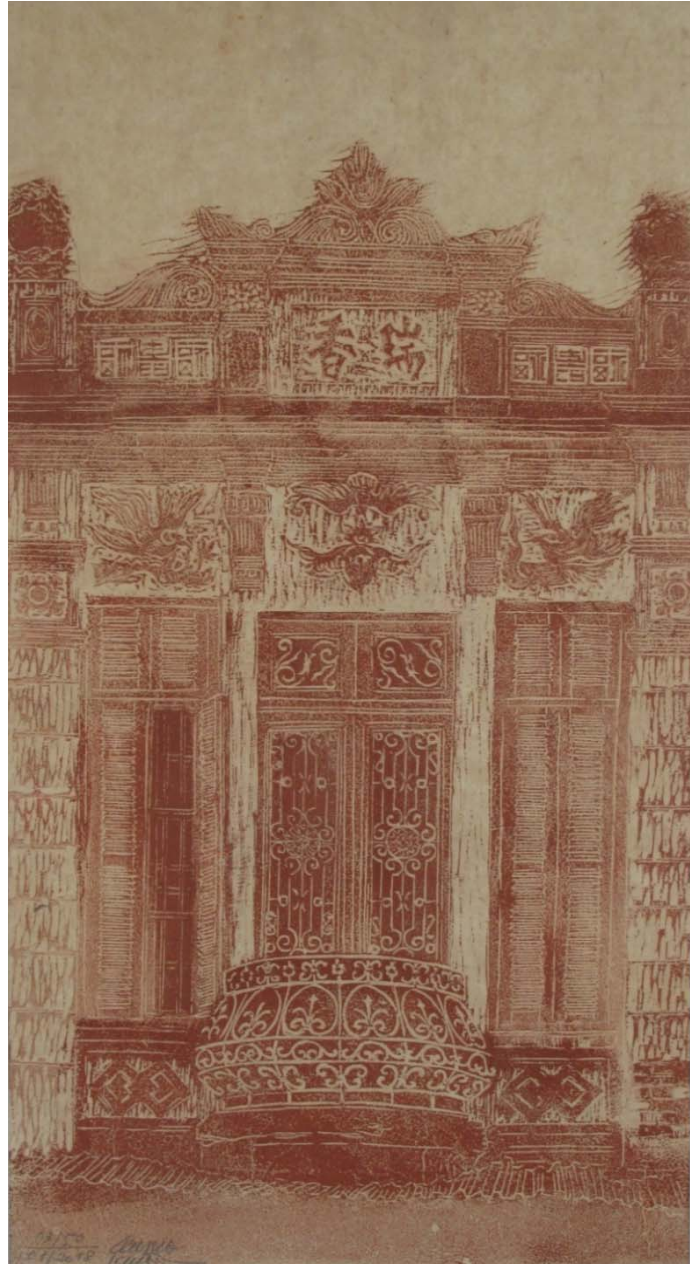

If Vietnamese fine art has intrigued myriad art collectors and researchers for its mixture of Southeast Asian artistic traditions since the early 20$^{th}$ century—such as the use of bright colors and dominant themes of farmland and countryside (Taylor, 2009)—and French colonial legacy (Vuong, Ho, et al., 2018), then its architecture has also captivated artists and ordinary people alike. Inspired by the spirit of Bui Xuan Phai and his immortalized paintings of Hanoi old streets, this study hopes to invite scholars of Vietnam studies as well as the international community to join the discussion on novel quantitative methodology in architecture studies and social sciences at large.

For closing this inquiry, the paper would like to end with a linocut painting by the artist Bui Quang Khiem (Figure 11). The picture, based on an actual photo taken by the artist for this research, shows the façade of an old house on Cau Go street in Hanoi that exemplifies the mixing of many features of different architecture styles, such as the prominent Chinese characters at the center of the pediment, the two lotus flowers on the top columns, the art décor window and balcony iron railings, the French shuttered windows, to name a few. Just as the house in this linocut print stands immortalized on paper, the 500 photos taken for this research project (of which 278 were used) have locked in the shape and form of Hanoi houses at a specific point in time. In a city that is bustling with constant movements, the photos freeze Hanoi and its unique fusion of foreign architectural elements in time just enough for this academic inquiry to set off.